\documentclass[conference]{IEEEtran}
\IEEEoverridecommandlockouts

\usepackage{amsmath,amssymb,amsfonts}
\usepackage{algorithmic}
\usepackage{textcomp}
\usepackage{xcolor}
\usepackage{lipsum}
\usepackage{graphicx}
\usepackage{multirow}
\usepackage{graphics}
\usepackage{subcaption}
\usepackage{scalerel}
\usepackage{verbatim}
\usepackage{hyperref}
\usepackage{caption}
\usepackage{subcaption}
\usepackage{float}
\usepackage{cite}

\def\BibTeX{{\rm B\kern-.05em{\sc i\kern-.025em b}\kern-.08em
    T\kern-.1667em\lower.7ex\hbox{E}\kern-.125emX}}
\begin{document}

\title{Adversarial Domain Adaptation with Paired Examples for Acoustic Scene Classification \\ on Different Recording Devices}

\author{\IEEEauthorblockN{Stanisław Kacprzak}
\IEEEauthorblockA{
\textit{AGH University of Science and Technology}\\
\textit{Institute of Electronics} \\
30-059 Krakow, Poland \\
skacprza@agh.edu.pl}
\and
\IEEEauthorblockN{Konrad Kowalczyk}
\IEEEauthorblockA{\textit{AGH University of Science and Technology}\\
\textit{Institute of Electronics} \\
30-059 Krakow, Poland \\
konrad.kowalczyk@agh.edu.pl}
\thanks{This research was supported by the National Science Centre under grant number DEC-2017/25/B/ST7/01792.}}

\maketitle

\begin{abstract}
In classification tasks, the classification accuracy diminishes when the data is gathered in different domains. To address this problem, in this paper, we investigate several adversarial models for domain adaptation (DA) and their effect on the acoustic scene classification task. The studied models include several types of generative adversarial networks (GAN), with different loss functions, and the so-called cycle GAN which consists of two interconnected GAN models. The experiments are performed on the DCASE20 challenge task 1A dataset, in which we can leverage the paired examples of data recorded using different devices, i.e., the source and target domain recordings. The results of performed experiments indicate that the best performing domain adaptation can be obtained using the cycle GAN, which achieves as much as 66\% relative improvement in accuracy for the target domain device, while only 6\% relative decrease in accuracy on the source domain. In addition, by utilizing the paired data examples, we are able to improve the overall accuracy over the model trained using larger unpaired data set, while decreasing the computational cost of the model training.
\end{abstract}


\section{Introduction}

Acoustic Scene Classification (ASC) is the task of assigning a predefined label to an audio segment that best describes its contents. The historical preview of previous research and general framework for ASC can be found in \cite{overview15}, and an overview of current methods based on deep learning can be found in \cite{deepOverview20}. The ASC is one of the main tasks of the DCASE challenge\footnote{\url{http://dcase.community/}}, and reviews of the most recent competitions can be found in \cite{dcase2018, Heittola2020_DCASE}.
In this paper, we focus on the specific problem of mismatched domains (also known as the domain shift), where the mismatch is primarily caused by the usage of different recording devices. The methods addressing this problem are known as domain adaptation (DA). In this work, we approach the domain adaptation problem via generative adversarial networks (GAN) \cite{goodfellow2014generative} and their extension CycleGAN \cite{zhu2017unpaired}. Previous use of GANs in the context of ASC include generation of new training examples \cite{mun2017generative} and extraction of features using deep convolutional GAN (DCGAN) \cite{arniriparian2018fusion}. A system with domain adaptation using GAN for the downsteam ASC task was first used in \cite{gharib2018unsupervised}. This method has been further improved using Wasserstein distance GAN (WGAN), with the underlying theoretical framework presented in \cite{drossos2019unsupervised}.
Other recent approaches of DA for the ASC task include spectrum correction \cite{nguyen2020acoustic}, band-wise statistics matching \cite{mezza2021unsupervised}, neural label embedding, relational teacher-student learning \cite{hu2020relational}, channel domain conversion \cite{mun2019domain}, and feature projection \cite{mezza2020feature}.
The theoretical foundations of domain adaptation can be found in \cite{ben2010theory}, and a general framework for adversarial domain adaptation is presented in \cite{tzeng2017adversarial}.

In this paper, we present several methods to perform adversarial domain adaptation, as pre-processing to the acoustic scene classification from the recordings of several different devices. In particular, we focus on GAN and cycle GAN models, with additional loss terms, including identity and transfer loss terms. During the training of adversarial models, we also take advantage of the availability of paired data, an aspect of the data set that not many studies exploit (by trying to solve a harder, but more common, task with unpaired data), and compare it to solutions based on unpaired training. In our experiments, we use the DCASE20 task 1A dataset, which consists of 9 different devices and 10 acoustic sound scenes, as opposed to 4 devices only used in previous DACSE18 and DCASE19 challenges. To asses the efficacy of DA, we perform experiments with an ASC classifier trained only on source data. Our goal is to investigate and subsequently identify the models which perform best in the DA task, rather than to find a solution which achieves the most accurate classification. A comparison with other approaches such as independent classification in each domain or application of transfer learning to the classifier will be subject to future research.

The paper is structured as follows. In Sec. \ref{sec:DA} we present the models with different loss functions for adversarial domain adaptation. Sec. \ref{sec:experiments}
presents the network architectures, data sets, training procedure, metrics. Results and discussion are presented in Sec. \ref{sec:results}, followed by a summary in Sec. \ref{sec:conclusions}.
\section{Domain adaptation with Generative Adversarial Networks}
\label{sec:DA}
In this section, we describe training objectives used in presented models for domain adaptation. Inspired by the use of the CycleGAN \cite{zhu2017unpaired} for domain adaptation in the context of speaker recognition \cite{cycledomains, nidadavolu2019low}, we propose to perform device characteristic translation of audio recordings made with different target devices to a single source device. We start with a simple generator trained on paired data of both domains, then we add a domain discriminator to form a GAN model, and finally, use CycleGAN consisting of two interconnected GAN models. 

Lets us first justify the decision to transform the input data to the source domain. Transforming source data to the target domain may seem to be an attractive solution since it is easier to degrade high quality audio of the source, rather than to reconstruct it. However, in the considered problem, the target domain contains several different devices, while the source domain consist of a single device, which makes domain adaptation to the source domain a well-defined transformation. In contrast, because of the existence of multiple devices in the target data set, the opposite transformation is ill-defined and would result in a transformation to some (most likely) non-existing device domain. 
\subsection{Generator model}
The generator  $G_{\scaleto{T \rightarrow S}{3pt}}$ is trained to learn the mapping from the  target domain $T$ to the source domain $S$. The training data $X_{\scaleto{S}{3pt}}$ and $X_{\scaleto{T}{3pt}}$ consists of elements from two separate distributions $x_{\scaleto{S}{3pt}} \sim p_{\scaleto{S}{3pt}}(x)$ and $x_{\scaleto{T}{3pt}} \sim p_{\scaleto{T}{3pt}}(x)$ and each $x_{\scaleto{S}{3pt}}$ has a corresponding (paired) $x_{\scaleto{T}{3pt}}$ element.
To obtain desired domain mapping the generator is trained using $\ell_1$ loss between the paired examples
\begin{equation}
\label{transfer_loss}
L_{T \rightarrow S}=\mathop{\mathbb{E}_{x_{\scaleto{T}{3pt}} \sim p_{\scaleto{T}{3pt}}}[||G_{\scaleto{T \rightarrow S}{3pt}}(x_{\scaleto{T}{3pt}}) - x_{\scaleto{S}{3pt}}||_1]}.
\end{equation}
However, since during the test time, the domain of the tested audio is unknown, the source audio should be invariant to the transformation. This results in an additional loss component
\begin{equation}
\label{identity}
L_{\mathrm{id}_{T \rightarrow S}}=\mathop{\mathbb{E}_{x_{\scaleto{S}{3pt}} \sim p_{\scaleto{S}{3pt}}}[||G_{\scaleto{T \rightarrow S}{3pt}}(x_{\scaleto{S}{3pt}}) - x_{\scaleto{S}{3pt}}||_1]},
\end{equation}
and thus the total loss function is defined as
\begin{equation}
\label{generator los}
L_{\mathrm{G}}^{\mathrm{total}} = L_{T \rightarrow S} + \lambda_{\mathrm{id}} \; L_{\mathrm{id}_{T \rightarrow S}},
\end{equation}
where $\lambda_{\mathrm{id}}$ is the weighting coefficient for the identity loss.
\subsection{GAN model}
In GAN approach, beside the $G_{\scaleto{T \rightarrow S}{3pt}}$ generator, there is also a domain discriminator
$D_S$, which is trained to recognize elements from domain $S$. We use the definition of GAN loss function using the mean square error (MSE)
\begin{align}
\label{gan_loss}
L_{\scaleto{\mathrm{GAN}_{T \rightarrow S}}{4pt}} =&\mathop{\mathbb{E}_{x_{\scaleto{T}{3pt}} \sim p_{\scaleto{T}{3pt}}}[D_{\scaleto{S}{3pt}}(G_{\scaleto{T \rightarrow S}{3pt}}(x_{\scaleto{T}{3pt}}))^2]} + \nonumber \\
&\mathop{\mathbb{E}_{x_{\scaleto{S}{3pt}} \sim p_{\scaleto{S}{3pt}}}[(D_{\scaleto{S}{3pt}}(x{\scaleto{S}{3pt}})-1)^2]}
\end{align}
and similarly to the generator loss, additional components can be added to the standard loss function. We propose to use the general loss function of the GAN which is formulated as
\begin{equation}
\label{eq:GAN_tot}
L_{\scaleto{\mathrm{GAN}_{T \rightarrow S}}{4pt}}^{\mathrm{total}} = L_{\scaleto{\mathrm{GAN}_{T \rightarrow S}}{4pt}} + \lambda_{\mathrm{tr}}  \;  L_{T \rightarrow S} + \lambda_{\mathrm{id}} \; L_{\mathrm{id}_{T \rightarrow S}},
\end{equation}
where $\lambda_{\mathrm{tr}}$ denotes the weighting coefficient for the transformation loss term. Similar loss (without $L_{\mathrm{id}_{T \rightarrow S}}$ component) was used in \cite{isola2018imagetoimage}.

\subsection{CycleGAN model}
Finally, the CycleGAN architecture consists of two interconnected GAN models. The first GAN is transforming the target domain to the source domain, and the second GAN is performing the opposite transformation. Note that although we are interested only in a one-way mapping, CycleGAN learns the mapping in both directions to allow regularization in the form of cycle-consistency, which requires reconstruction of the original features with minimum error. The cycle consistency loss is defined as 
\begin{align}
\label{cycle_loss}
L_{\mathrm{Cyc}} =&\mathop{\mathbb{E}_{x_{\scaleto{S}{3pt}} \sim p_{\scaleto{S}{3pt}}}[||G_{\scaleto{T \rightarrow S}{3pt}}(G_{\scaleto{S \rightarrow T}{3pt}}(x_{\scaleto{S}{3pt}}))-x_{\scaleto{S}{3pt}}||_1]} + \nonumber\\
&\mathop{\mathbb{E}_{x_{\scaleto{T}{3pt}} \sim p_{\scaleto{T}{3pt}}}[||G_{\scaleto{S \rightarrow T}{3pt}}(G_{\scaleto{T \rightarrow S}{3pt}}(x_{\scaleto{T}{3pt}}))-x_{\scaleto{T}{3pt}}||_1]}\,.
\end{align}
Finally, we use the total loss for the proposed CycleGAN which is defined as
\begin{align}
\label{total}
L_{\scaleto{\mathrm{CycGAN}}{4pt}}^{\mathrm{total}}=&L_{\scaleto{\mathrm{GAN}_{S \rightarrow T}}{4pt}} + L_{\scaleto{\mathrm{GAN}_{T \rightarrow S}}{4pt}} + \lambda_{\mathrm{cyc}} \;  L_{{\scaleto{\mathrm{cyc}}{4pt}}} \nonumber\\
+& \lambda_{\mathrm{id}} \; (L_{\mathrm{id}_{S \rightarrow T}} + L_{\mathrm{id}_{T \rightarrow S}})\,,
\end{align}
where $\lambda_{\mathrm{cyc}}$ denotes the weighting coefficient for the cycle consistency loss term.
\section{Experiments and Evaluation}
\label{sec:experiments}
\subsection{Network architectures for domain adaptation}
The network architecture of the evaluated CycleGAN model largely follows \cite{cycledomains}. The generator network consists of the downsampling and upsampling blocks. We choose the best performing version with a skip connection that adds downsampler input to the upsampler output, which allows to preserve the input structure and forces the generator to learn the differences between the source and the target domains. Our modification to the architecture consists in removing the ReLU activation function at the end of the residual block, which is in line with the implementation of the original CycleGAN \cite{zhu2017unpaired} and is beneficial according to\footnote{\url{http://torch.ch/blog/2016/02/04/resnets.html}}. We also add a non-linear activation of outputs \emph{tanh} to the generator (similarly to the implementation of \cite{zhu2017unpaired}), and apply a \emph{sigmoid} activation of outputs in the discriminator, since we found in the preliminary experiments that this makes the training more stable.
Note that in this work, the same generator architecture is used in all compared models including GAN and the presented CycleGAN. As a result, the generators of the trained models, which are later used in pre-processing of the ASC task, differ only in the estimated network parameters.  

The implementation of all models is done with the Pytorch Lightning framework \cite{falcon2019pytorch}.

\subsection{Data set and preprocessing}
\label{data}
The data set used in the experimental evaluation is a development set of the TAU Urban Acoustic Scenes 2020 Mobile data set \cite{Mesaros2018_DCASE}. The development set contains data from 10 European cities and 10 acoustic scenes: airport, shopping mall, metro station, pedestrian street, public square, a street with a medium level of traffic, traveling by tram, traveling by bus, traveling by an underground metro, and urban park. Recordings are made with 9 different devices, from which 3 are real devices (A, B, C) and 6 are simulated devices (S1-S6). The total amount of audio in the development set amounts to 64 hours and it consists of 10-second long recordings. Most of the recordings (40h duration) are made with device A, which is the high-quality equipment recording at a sampling rate of 48kHz and with a 24-bit resolution, to which we will refer to as a \emph{source}. Other recordings (24h duration) are made using the devices of lower audio quality, i.e., a mobile phone, a GoPro camera, and several simulated devices obtained by using simulated impulse responses (IRs) and dynamic range control, to which we will refer to as the \emph{target} recordings.

For every recording of the target device, there is an overlapping (paired) recording with the source device, but there are recordings of the source device without the corresponding target device recording. For that reason, our paired data set contains only 19h of audio recordings as opposed to 39h in unpaired data set. To be able to provide reliable results without an access to the DCASE evaluation data set, we extract separate validation set from the test split. Importantly, the test set contains two devices that are not present in the training and validation part (S5 and S6). As advised in the challenge description, recordings with the same location ID are allowed only in one split (for that reason some recordings of the new devices are excluded from the new test split). Finally, our validation set consists of 7 devices and about 100 recordings per device such that the test set consists of 9 devices with around 230 recordings per device.

Regarding the input features to the evaluated networks, the data is transformed to the short-time Fourier (STFT) domain, then a log-Mel filter bank is applied to the power in the STFT domain as in \cite{Hu2020} except that we use a lower number of frequency bands, which amounts to 40 (as in \cite{cycledomains}). The final feature map is computed by taking the logarithm of the power spectogram and rescaling it to the values in the range [-1, 1].

\subsection{Acoustic scene classification}
To evaluate the impact of the proposed domain adaptation on acoustic scene classification (ASC), we choose a ResNet-based classifier which is described in \cite{Hu2020}, with the code shared by the authors\footnote{\url{https://github.com/MihawkHu/DCASE2020_task1}}. We choose the simpler version of the classifier without any data augmentation techniques and with a 40-band log-Mel filter bank along with delta and delta-delta features scaled to the range [0, 1]. Nonetheless, note that the applied system achieves better results than the provided DCASE baseline classifier\footnote{\url{https://github.com/toni-heittola/dcase2020_task1_baseline}}.

\subsection{Training and testing}
Beside using the paired data, we performed also experiments with training using all available training data (using all recordings as unpaired data), in which we make sure that both source and target recordings always represent the same acoustic scene.
The ASC classifier is trained for 200 epochs with the configuration provided by the authors, except for changing the input dimensions to match our generator's output and monitoring the classifier accuracy with our validation set. The ASC classifier is trained using only the source data.
All of our generative models use training hyperparameters similar to the ones used in \cite{tzeng2017adversarial} and \cite{cycledomains}. The model input is built of fragments of Mel-spectrograms consisting of 11 contiguous frames processed in batches of size 32. Adam optimizer is used with $\beta_1=0.5$ and $\beta_2=0.999$. The learning rate is set to $0.002$ with a linear decay after 15 epochs. Loss weight  $\lambda_{\mathrm{cyc}}$ is set to 10, $\lambda_{\mathrm{id}}$ set to 1 or 5  and $\lambda_{\mathrm{tr}}$ is set to 5 or 0. When training generators in practice we use the mean rather than the sum in \eqref{total}, dividing the loss by two.
The models are trained for no more than 200 epochs and their performance is measured every third epoch as an accuracy obtained by the ASC classifier on the validation set after domain adaptation using the trained model.
\subsection{Evaluated systems}
In performed experiments, in total we evaluate 6 models for domain adaptation: (i) the generator with loss given by \eqref{generator los} with $\lambda_{\mathrm{id}}$ = 1 denoted as $\mathrm{Generator}$, (ii) identity preserving GAN given by \eqref{eq:GAN_tot} with $\lambda_{\mathrm{id}}$ = 5 and $\lambda_{\mathrm{tr}}$ = 0, denoted as $\mathrm{GAN_{id}}$, (iii) GAN with identity and transfer loss functions given by \eqref{eq:GAN_tot} with $\lambda_{\mathrm{id}}$ = 5 and $\lambda_{\mathrm{tr}}$ = 5, denoted as $\mathrm{GAN_{id,tr}}$, (iv) GAN trained using unpaired data with an identity loss given by \eqref{eq:GAN_tot} with $\lambda_{\mathrm{id}}$ = 5 and $\lambda_{\mathrm{tr}}$ = 0, denoted as $\mathrm{GAN_{id}^{all}}$, (v) cycle GAN with the loss given by \eqref{total} with $\lambda_{\mathrm{id}}$ = 5 and $\lambda_{\mathrm{cyc}}$ = 10, denoted as $\mathrm{CycleGAN}$, and finally (vi) cycle GAN trained using unpaired data with $\lambda_{\mathrm{id}}$ = 5 and $\lambda_{\mathrm{cyc}}$ = 10, denoted as $\mathrm{CycleGAN^{all}}$. For comparison, we also evaluate the system without domain adaptation, which is denoted as NA.  
\subsection{Evaluation measures}
To evaluate the impact of applying the investigated DA methods on acoustic scene classification, we measure ASC accuracy using a classifier trained using only the source data. During the final prediction all test examples are transformed to the source domain using the studied DA systems.
To be able to directly measure how DA affects the data, we also measure the Log-Spectral Distance (LSD) \cite{rabiner1993fundamentals} between the paired source and target domain examples (to be able to calculate those distances, we used also the data from source domain that was not included in the cross-validation split). Since we use the scaled log-Mel-spectograms during the network training, for consistency we apply the analogous scaling to the log-Mel-spectograms when computing the LSDs, which as a result reduces those values to the range from 0 to 10 dB.
\section{Results and discussion}

\label{sec:results}
\begin{table}[!t]
	\centering
	\caption{Accuracy of the ASC classifier for different models performing domain adaptation tested on the source domain data, target domain data, and only on devices unseen during the model training (denoted as new devices).}
	\label{tab:accuracy}
	\begin{tabular}{lccc}
		DA method & \shortstack{Accuracy [\%] \\ (source)} & \shortstack{Accuracy [\%] \\ (target)} & \shortstack{Accuracy [\%] \\ (new devices)} \\
		\hline
		\hline
		NA & 74.89 & 25.06 & 15.55\\
		\hline
		$\mathrm{Generator}$ & 70.04 & 19.23 & 19.87\\
		\hline
		$\mathrm{GAN_{id}}$ & \textbf{75.33} & 25.88 & 15.76 \\
		$\mathrm{GAN_{id, tr}}$ & 63.00 & 40.19 & \textbf{28.51} \\
		$\mathrm{GAN_{id}^{all}}$ & 75.33 & 25.39 & 16.84\\
		\hline
		$\mathrm{CycleGAN}$ & 70.48 & \textbf{41.56} & 24.62 \\
		$\mathrm{CycleGAN}^{\mathrm{all}}$& 70.04 & 39.87 & 27.00\\
	\end{tabular}
\end{table}
\begin{table}[!t]
	\centering
	\caption{Mean log-spectral distance (LSD) between the source and target domain recordings tested on the source domain data, target domain data, and only on devices unseen during the model training (denoted as new devices).}
	\label{tab:lsd}
	\begin{tabular}{lccc}
		DA method & \shortstack{LSD [dB] \\ (source)}    & \shortstack{LSD [dB] \\ (target)} & \shortstack{LSD [dB] \\ (new devices)} \\
		\hline
		\hline
		NA  &  0.000 & 1.399 & 1.668\\
		\hline
		$\mathrm{Generator}$ & 0.073 & \textbf{0.723} & 1.188\\
		\hline
		$\mathrm{GAN_{id}}$ & 0.055 & 1.375 & 1.640\\
		$\mathrm{GAN_{id, tr}}$  & 0.120 & 0.766 & 1.216\\
		$\mathrm{GAN_{id}^{all}}$  & 0.054 & 1.387 & 1.650\\
		\hline
		$\mathrm{CycleGAN}$ & 0.082 & 0.813 & 1.138 \\
		$\mathrm{CycleGAN^{all}}$ & 0.081 & 0.790 & \textbf{1.062} \\
	\end{tabular}
\end{table}
Table \ref{tab:accuracy} presents the results of ASC accuracy obtained on test data transformed to the source domain using the proposed systems. We analyze accuracy separately for the data from the source device (A), all other devices (B,C, S1-S6), and devices not presented in training and validation splits (S5 and S6). Note that there is eight times more target devices than source devices in the test set (see section \ref{data}). We also calculate the mean log-spectral distance (LSD) for the domain adapted data on the same 3 subsets of the test data set, those results are presented in Table \ref{tab:lsd}. We can observe that the separately trained generator is able to obtain the smallest LSD value for target devices. However, this does not lead to the overall best performance in terms of the classification accuracy, which is the worst among all 6 compared DA models, with an improvement observed only for new devices. The GAN model provides a steady, small, yet very consistent increase in acoustic scene classification accuracy across all 3 subsets of the test data. In particular, a large increase in the accuracy for target devices is observed when the transfer loss function is additionally incorporated into the GAN, however, this comes at a cost of reducing the accuracy for the source data. This improvement could be attributed to the discriminator that enforces generated outputs to be similar to source domain and prevents finding poor local minima of (\ref{generator los}). The best overall performance can be observed for the CycleGAN models, which achieve high accuracy for the target devices, while they exhibit only a relatively small decrease in accuracy for the source devices. As the most general, it takes advantages of the aforementioned models and provides good generalization to data from new, unseen devices. Finally, CycleGAN trained using unpaired data is able to achieve high accuracy and low LSD for new devices. Its performance on target devices is close to GAN with transfer loss, but performs better for source devices, probably thanks to having more data in the unpaired data set. However, overall the better accuracy is obtained when using CycleGAN trained with paired data. Note that using paired data results in a smaller training data set, which also allows for a faster training time.
\begin{figure}[!t]
\label{figure}
    \centering
    \includegraphics[width=0.82\columnwidth]{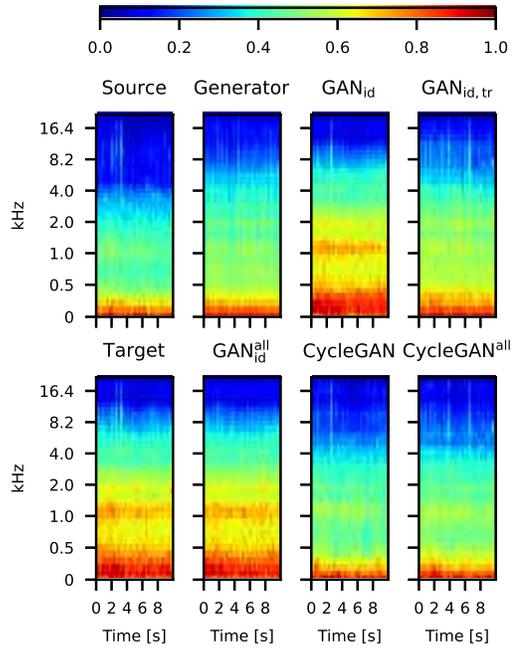}
	\caption{Normalized log-Mel-Spectograms of the source and target recordings with and without the domain adaptation using the investigated models for the example files \emph{park-paris-2424-7265-a.wav} and \emph{park-paris-2424-7265-s5.wav} from the TAU Urban Acoustic Scenes 2020 Mobile data set \cite{Mesaros2018_DCASE}.}
	\label{fig:fig1}
\end{figure}
In order to provide additional insights into the differences between the compared models for domain adaption, we present normalized log-Mel-spectograms before and after domain adaptation for an example audio file. As can be seen from Fig. \ref{fig:fig1}, DA using any of the two CycleGANs clearly provides the best match with the source domain, with the GAN with identity and transfer loss functions and the pure generator also quite successfully performing domain adaptation. The final result is shown in Fig. \ref{fig:tsne}, in which we present the 2D data visualization for two acoustic scenes recorded using four different types of devices, before and after domain adaptation using the best performing CycleGAN. As can be observed, the data before DA is clearly clustered, with clusters corresponding to the devices and sound scenes. After domain adaptation, the clusters corresponding to the different devices are not easily separable anymore, which indicates that device-related domain adaptation is successfully performed.
\begin{figure}[!t]
    \centering
    \includegraphics[width=0.5\textwidth]{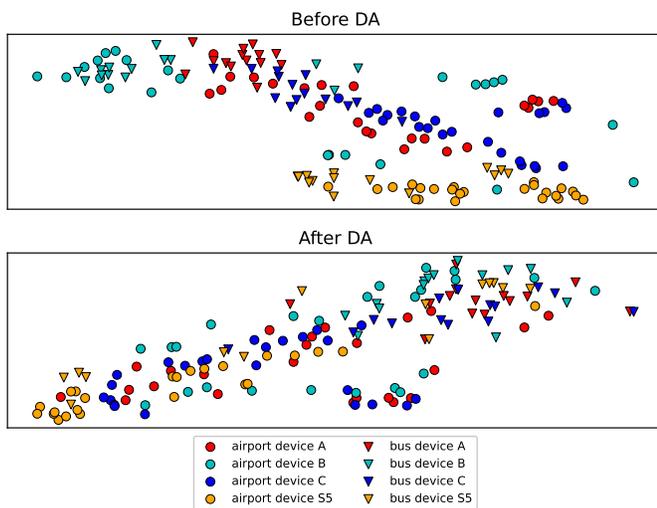}
	\caption{T-SNE \cite{van2008visualizing} visualization of domain adaptation effects on the \emph{airport} and \emph{bus} test data recorded using four types of devices, i.e., source device A and target devices B, C, and S5. The top plot shows the data before DA and the bottom plot shows the data after domain adaptation using the Cycle GAN.}
	\label{fig:tsne}
\end{figure}

\section{Conclusions}
\label{sec:conclusions}
In this paper, we analyze the impact of domain adaptation using six presented models on the acoustic scene classification task. In particular, we compare the domain adaptation achieved by the simple generator, three types of the GAN model, and a CycleGAN, paying special attention to the impact of training these models using paired and unpaired data. The results of performed experiments show that the highest accuracy is obtained using the CycleGAN, trained also with the paired data as provided in the TAU Urban Acoustic Scenes 2020 Mobile data set. The best-performing domain adaptation system achieved 66\% of relative improvement in accuracy for the data recorded using the target devices, with 6\% relative decrease in accuracy on the data recorded using the source device.
\bibliographystyle{IEEEtran}
\bibliography{references.bib}

\begin{thebibliography}{10}
\providecommand{\url}[1]{#1}
\csname url@samestyle\endcsname
\providecommand{\newblock}{\relax}
\providecommand{\bibinfo}[2]{#2}
\providecommand{\BIBentrySTDinterwordspacing}{\spaceskip=0pt\relax}
\providecommand{\BIBentryALTinterwordstretchfactor}{4}
\providecommand{\BIBentryALTinterwordspacing}{\spaceskip=\fontdimen2\font plus
\BIBentryALTinterwordstretchfactor\fontdimen3\font minus
  \fontdimen4\font\relax}
\providecommand{\BIBforeignlanguage}[2]{{%
\expandafter\ifx\csname l@#1\endcsname\relax
\typeout{** WARNING: IEEEtran.bst: No hyphenation pattern has been}%
\typeout{** loaded for the language `#1'. Using the pattern for}%
\typeout{** the default language instead.}%
\else
\language=\csname l@#1\endcsname
\fi
#2}}
\providecommand{\BIBdecl}{\relax}
\BIBdecl

\bibitem{overview15}
D.~{Barchiesi}, D.~{Giannoulis}, D.~{Stowell}, and M.~D. {Plumbley}, ``Acoustic
  scene classification: Classifying environments from the sounds they
  produce,'' \emph{IEEE Signal Processing Magazine}, vol.~32, no.~3, pp.
  16--34, 2015.

\bibitem{deepOverview20}
J.~Abeßer, ``A review of deep learning based methods for acoustic scene
  classification,'' \emph{Applied Sciences}, vol.~10, no.~6, 2020.

\bibitem{dcase2018}
S.~{Gharib}, H.~{Derrar}, D.~{Niizumi}, T.~{Senttula}, J.~{Tommola},
  T.~{Heittola}, T.~{Virtanen}, and H.~{Huttunen}, ``Acoustic scene
  classification: A competition review,'' in \emph{2018 IEEE 28th International
  Workshop on Machine Learning for Signal Processing (MLSP)}, 2018, pp. 1--6.

\bibitem{Heittola2020_DCASE}
T.~Heittola, A.~Mesaros, and T.~Virtanen, ``Acoustic scene classification in
  dcase 2020 challenge: generalization across devices and low complexity
  solutions,'' in \emph{Proceedings of the Detection and Classification of
  Acoustic Scenes and Events 2020 Workshop (DCASE2020)}, Tokyo, Japan, November
  2020, pp. 56--60.

\bibitem{goodfellow2014generative}
I.~Goodfellow \emph{et~al.}, ``Generative adversarial nets,'' ser.
  NIPS'14.\hskip 1em plus 0.5em minus 0.4em\relax Cambridge, MA, USA: MIT
  Press, 2014, p. 2672–2680.

\bibitem{zhu2017unpaired}
J.-Y. Zhu, T.~Park, P.~Isola, and A.~A. Efros, ``Unpaired image-to-image
  translation using cycle-consistent adversarial networks,'' in
  \emph{Proceedings of the IEEE International Conference on Computer Vision},
  2017, pp. 2223--2232.

\bibitem{mun2017generative}
S.~Mun, S.~Park, D.~K. Han, and H.~Ko, ``Generative adversarial network based
  acoustic scene training set augmentation and selection using {SVM}
  hyper-plane,'' \emph{Proc. DCASE}, pp. 93--97, 2017.

\bibitem{arniriparian2018fusion}
S.~Arniriparian, M.~Freitag, N.~Cummins, M.~Gerczuk, S.~Pugachevskiy, and
  B.~Schuller, ``A fusion of deep convolutional generative adversarial networks
  and sequence to sequence autoencoders for acoustic scene classification,'' in
  \emph{2018 26th European Signal Processing Conference (EUSIPCO)}, 2018, pp.
  977--981.

\bibitem{gharib2018unsupervised}
S.~Gharib, K.~Drossos, E.~Cakir, D.~Serdyuk, and T.~Virtanen, ``Unsupervised
  adversarial domain adaptation for acoustic scene classification,'' in
  \emph{Proceedings of the Detection and Classification of Acoustic Scenes and
  Events 2018 Workshop (DCASE2018)}, November 2018, pp. 138--142.

\bibitem{drossos2019unsupervised}
K.~Drossos, P.~Magron, and T.~Virtanen, ``Unsupervised adversarial domain
  adaptation based on the {W}asserstein distance for acoustic scene
  classification,'' in \emph{IEEE Workshop on Applications of Signal Processing
  to Audio and Acoustics (WASPAA)}, 2019, pp. 259--263.

\bibitem{nguyen2020acoustic}
T.~Nguyen, F.~Pernkopf, and M.~Kosmider, ``Acoustic scene classification for
  mismatched recording devices using heated-up softmax and spectrum
  correction,'' in \emph{IEEE International Conference on Acoustics, Speech and
  Signal Processing (ICASSP)}, 2020, pp. 126--130.

\bibitem{mezza2021unsupervised}
A.~I. Mezza, E.~A. Habets, M.~M{\"u}ller, and A.~Sarti, ``Unsupervised domain
  adaptation for acoustic scene classification using band-wise statistics
  matching,'' in \emph{2020 28th European Signal Processing Conference
  (EUSIPCO)}, 2021, pp. 11--15.

\bibitem{hu2020relational}
H.~Hu, S.~M. Siniscalchi, Y.~Wang, and C.-H. Lee, ``Relational teacher student
  learning with neural label embedding for device adaptation in acoustic scene
  classification,'' \emph{arXiv preprint arXiv:2008.00110}, 2020.

\bibitem{mun2019domain}
S.~Mun and S.~Shon, ``Domain mismatch robust acoustic scene classification
  using channel information conversion,'' in \emph{IEEE International
  Conference on Acoustics, Speech and Signal Processing (ICASSP)}, 2019, pp.
  845--849.

\bibitem{mezza2020feature}
A.~I. Mezza, E.~A. Habets, M.~M{\"u}ller, and A.~Sarti, ``Feature
  projection-based unsupervised domain adaptation for acoustic scene
  classification,'' in \emph{2020 IEEE 30th International Workshop on Machine
  Learning for Signal Processing (MLSP)}, 2020, pp. 1--6.

\bibitem{ben2010theory}
S.~Ben-David, J.~Blitzer, K.~Crammer, A.~Kulesza, F.~Pereira, and J.~W.
  Vaughan, ``A theory of learning from different domains,'' \emph{Machine
  learning}, vol.~79, no.~1, pp. 151--175, 2010.

\bibitem{tzeng2017adversarial}
E.~Tzeng, J.~Hoffman, K.~Saenko, and T.~Darrell, ``Adversarial discriminative
  domain adaptation,'' in \emph{Proceedings of the IEEE Conference on Computer
  Vision and Pattern Recognition}, 2017, pp. 7167--7176.

\bibitem{cycledomains}
P.~S. Nidadavolu, J.~Villalba, and N.~Dehak, ``{Cycle-GANs} for domain
  adaptation of acoustic features for speaker recognition,'' in \emph{IEEE
  International Conference on Acoustics, Speech and Signal Processing
  (ICASSP)}, 2019, pp. 6206--6210.

\bibitem{nidadavolu2019low}
P.~S. Nidadavolu, S.~Kataria, J.~Villalba, and N.~Dehak, ``Low-resource domain
  adaptation for speaker recognition using {Cycle-GANs},'' in \emph{2019 IEEE
  Automatic Speech Recognition and Understanding Workshop (ASRU)}, 2019, pp.
  710--717.

\bibitem{isola2018imagetoimage}
P.~Isola, J.-Y. Zhu, T.~Zhou, and A.~A. Efros, ``Image-to-image translation
  with conditional adversarial networks,'' in \emph{Proceedings of the IEEE
  Conference on Computer Vision and Pattern Recognition}, 2017, pp. 1125--1134.

\bibitem{falcon2019pytorch}
W.~Falcon \emph{et~al.}, ``Pytorch lightning,''
  \url{https://github.com/PyTorchLightning/pytorch-lightning}.

\bibitem{Mesaros2018_DCASE}
A.~Mesaros, T.~Heittola, and T.~Virtanen, ``A multi-device dataset for urban
  acoustic scene classification,'' in \emph{Proceedings of the Detection and
  Classification of Acoustic Scenes and Events 2018 Workshop (DCASE2018)},
  November 2018, pp. 9--13.

\bibitem{Hu2020}
H.~Hu \emph{et~al.}, ``A two-stage approach to device-robust acoustic scene
  classification,'' in \emph{IEEE International Conference on Acoustics, Speech
  and Signal Processing (ICASSP)}, 2021, pp. 845--849.

\bibitem{rabiner1993fundamentals}
L.~Rabiner, \emph{Fundamentals of speech recognition}.\hskip 1em plus 0.5em
  minus 0.4em\relax PTR Prentice Hall, 1993.

\bibitem{van2008visualizing}
L.~Van~der Maaten and G.~Hinton, ``Visualizing data using {t-SNE},''
  \emph{Journal of machine learning research}, vol.~9, no.~11, 2008.

\end{thebibliography}
\end{document}